# CHECKPOINTING WITH MINIMAL RECOVERY IN ADHOCNET BASED TMR

Sarmistha Neogy

Department of Computer Science & Engineering, Jadavpur University, India

## *Abstract:*

*This paper describes two-fold approach towards utilizing Triple Modular Redundancy (TMR) in Wireless Adhoc Network (AdocNet). A distributed checkpointing and recovery protocol is proposed. The protocol eliminates useless checkpoints and helps in selecting only dependent processes in the concerned checkpointing interval, to recover. A process starts recovery from its last checkpoint only if it finds that it is dependent (directly or indirectly) on the faulty process. The recovery protocol also prevents the occurrence of missing or orphan messages. In AdocNet, a set of three nodes (connected to each other) is considered to form a TMR set, being designated as main, primary and secondary. A main node in one set may serve as primary or secondary in another. Computation is not triplicated, but checkpoint by main is duplicated in its primary so that primary can continue if main fails. Checkpoint by primary is then duplicated in secondary if primary fails too.*

## 1. INTRODUCTION

Distributed systems that execute processes on different nodes connected by a communication network [6] are prone to failure. One of the widely used approaches for providing fault tolerance is the checkpoint/rollback recovery mechanism. Checkpointing is the method of periodically recording the state of the system in stable storage. The saving of process state information may be required for error recovery, debugging and other distributed applications [7]. This periodically saved state is called the *checkpoint* of the process [7, 8]. A global state [22] is a set of individual process states, one per process [7]. The state contains a snapshot at some instant during the execution of a process. The snapshot is required to be consistent to avoid the domino effect [23] that is, multiple rollbacks during recovery.

One of the most well-known methods of achieving fault tolerance is Triple Modular Redundant (TMR) [25] system. A minimum of three processors also known as replicas form a redundant group to perform replicated processing. Identical processing and distributed voting are performed on same input data. Intermediate result or output from each replica is exchanged among each other and majority voted upon. After successful majority voting the replicas either resume processing on the intermediate results or end their computation if it had been the final result. Communications among the replicas take place via communication links. Replica at the receiving end has to wait for a time-out period [26] for receiving before concluding that there may be fault in other part of the system. This concept of TMR is utilized in this work as a measure for achieving fault tolerance in a wireless adhoc network (AdhocNet) where a group of three nodes, known as mobile hosts (MH) form the three replicas. Also fault tolerance may be achieved by periodically using stable storage of the MHs to save the process' states, better known as checkpoints, during failure-free execution. When a failure occurs, the failed process restarts from its latest checkpoint. This minimizes the amount of lost computation. The proposed system is recoverable even if more than one failure (at most two) occurs in a TMR node. As is well known wireless adhoc network does not have any infrastructure facilities and hence each MH acts as router also. The concept of distributed systems is extended to the wireless environment.

A TMR group consists of MHs that act as main, primary and secondary. The TMR groups are not exclusive, that is, MH acting as main may act as primary in another TMR group and so on. The concept of TMR is modified here in the sense that the three MHs do not perform identical processing throughout. But the checkpoint taken by a main MH is replicated in its primary MH. This is because in case a main MH fails, the primary MH can continue from the latest checkpoint. Similarly, the secondary MH receives a copy of the checkpoint every time the primary MH takes one. This continues until the primary MH fails. The communicating partners of this TMR group however are unaware of this change in the actual partner at the other end. It is assumed that several processes that are running on the MHs may communicate with each other depending on application requirement.

In the present work checkpointing is initiated by a process of the system. In fact, each of the processes takes turn to act as the initiator. Generally, processes take local checkpoints *after being notified* by the initiator excepting special cases described in later sections. The processes synchronize their activities of the current checkpointing interval before finally committing their checkpoints. This removes inconsistency, if any, and then checkpoints are committed. The technique adopted in the present paper thus disallows the formation of neither zigzag paths nor zigzag cycle [4,11]. The checkpointing pattern described in the present paper takes only those checkpoints that will contribute to a consistent global snapshot thereby eliminating the number of "useless" (checkpoints that do not contribute to global consistency) checkpoints. Maintaining consistency is necessary to avoid the domino effect in case any process fails after taking its ith checkpoint. If the set of the ith checkpoints can be proved to be consistent, then in case of recovery the system has to roll back only up to the ith checkpoint since that set provides a consistent and hence a stable global state of the system.

The processes do not append status information with each and every computation message but keeps updating own status whenever a message is sent or received. This information is required to find process dependence during recovery. Though the simplest way is to roll back all processes but this makes some unnecessary rollbacks. To avoid such rollbacks the processes in the present system exchange status information whenever a rollback is decided. Each process can find out for itself whether it requires rollback depending upon its relationship with the failed process.

This paper describes that any global checkpoint taken in the above-mentioned fashion in the present system is not only consistent but also eliminates taking unnecessary checkpoints and the system has to roll back only to the last saved state in case of a failure. Also all processes in the system do not have to rollback following the rollback algorithm described in the paper. The rest of the paper is organized as follows. Section 2 throws light on some related works in this area.

Section 3 describes the system model, Section 4 discusses in details the checkpointing algorithm with a proof, Section 5 discusses the rollback procedure and the algorithm, Section 6 discusses integration of the activities of TMR AdhocNet and the last one, that is, Section 7 concludes the paper.

## 2. RELATED WORKS

With reference to Chandy & Lamport [1] and Wang et. al [24] Tsai & Kuo [23] states that "A global checkpoint M is consistent if no message is sent after a checkpoint of M and received before another checkpoint of M". Following these observations we regard consistency as the scenario where if a sender 'S' sends a message 'm' *before* it has taken its ith checkpoint, then message 'm' must have to be received by a receiver 'R' *before* the receiver has taken its ith checkpoint. A message will be termed *missing* if its send is recorded but receipt is not and otherwise it is termed as *orphan* [21]. Suppose a node fails after taking its ith checkpoint. It is desirable that the system in such a scenario should roll back to the last (ith) saved state and resume execution from there. If a system can ensure that there is no *missing* or *orphan* message in the concerned ith global checkpoint, then the set of all the ith checkpoints taken by its constituent processes is bound to be *consistent*. Unlike the approach that should exist in a distributed system Kalaiselvi and Rajaraman [5] have kept record at the message sending end and at the message receiving end and a checkpoint coordinator matches the log it gets from all the processes at each checkpointing time. The present system also keeps records of messages sent and received in each process but the log is matched in a distributed fashion. Due to disparity in speed or congestion in the network, a message belonging to (i+1)th checkpointing interval may reach its receiver who has not yet taken its ith checkpoint. Such a message is discarded in [21] and sender retransmits it. Another method of dealing with such messages is to prevent their occurrences by compelling the sender to wait for a certain time before sending a message after any checkpoint [13]. The present work discards such a message by adopting a technique in receiving whereas in another approach [11] any process refrains from sending during the interval between the receipt of checkpoint initiation message and completion of committing that checkpoint. Distributed systems that use the *recovery block* approach [17] and have a common time base may estimate a time by which the participating processes would take *acceptance tests*. These estimated instants form the *pseudo recovery point* times as described in [16]. The disadvantages of such a scheme are more than one, like, fast processes may have to wait for slow processes to catch up and other fault tolerance mechanisms like time out may be required. In [9,10] the authors have analyzed checkpoints taken in a distributed system having loosely synchronized clocks [13,14,18,19]. No special synchronization messages have been used in those methods but the existing clock synchronization messages were utilized. The work described in [4] however, allows processes to take checkpoints on one's own and then a consistent global checkpoint is constructed from the set of local checkpoints. The drawback of the method is that useless checkpoints can not be avoided. The approach taken by Strom et al. in [20] does not maintain a consistent global checkpoint at all times but has to save enough information to construct such a checkpoint when need arises. So, this requires logging of messages. Contrary to the present checkpointing protocol, The authors in [2,15] presents minimal snapshot collection protocol where dependency is calculated during checkpointing also and hence the actual time taken for formal commitment or abort of a checkpoint is not fixed. The concept of weight distribution and collection by the initiator in [2,15] appears superfluous and can be replaced if a participating process sends a list of processes dependent on it to the initiator. The overhead of checkpointing (in terms of the number of checkpoints) is great in all of the CAS, CBR, CASBR and NRAS [8] protocols in comparison to the protocol presented in this paper. The present protocol possesses the Rollback-Dependency-Trackability (RDT) property as described in [22] as shown in Section 5.

A few checkpointing recovery techniques for mobile computing systems (infrastructured wireless and mobile networks) are described in the literature. In the two-tier checkpointing approach [27], coordinated checkpointing is used between the Mobile Service Stations (MSS) to reduce the number of checkpoints to be stored to a minimum. In [28] the log of unacknowledged messages are kept at stable storage of the home station (that is an MSS) of the mobile host. Gass and Gupta [29] in their algorithm takes three kinds of checkpoints—communication induced (taken after receiving an application message), local checkpoint (when an MH leaves the MSS to which it is connected to) and forced checkpoints (only the local variables are updated). All applications are assumed to be blocked during the algorithm execution thereby wasting computation power and information that failure has occurred is assumed to reach all fault free processes within finite time which is difficult in reality. This algorithm saves battery power by minimizing the recomputation time.

The work in [30] describes checkpointing and recovery using TMR in wireless infrastructured network. The authors have described a checkpointing and recovery protocol for infrastructured mobile system in [31]. The authors in [32] utilized the concept of mobile agents for checkpointing purposes in mobile systems. The works in [31] and [32] utilized different approaches towards checkpointing for infrastructure mobile systems. The authors have considered an attack model and augmented Mobile Adhoc Network with security features in their work in [33]. The work in [33] has enabled us to consider any AdhocNet routing algorithm for the present work.

## 3. SYSTEM MODEL AND ASSUMPTIONS

Let us consider a system of 'n' processes, $P_0$, $P_1$, …. $P_{n-1}$. Let the checkpoints (for the kth process) be denoted as the initial checkpoint $CP_k^0$ (i = 0), first checkpoint $CP_k^1$ (i=1), second checkpoint $CP_k^2$ (i=2) and so on. The time interval between any two consecutive checkpoints is called checkpointing interval that is eventually the next checkpoint number. This means that, the first checkpointing interval is the interval between the initial checkpoint and the first checkpoint. The initial checkpoint is taken when the system is being initialized. The processes communicate via messages only. We assume the following properties of the system:

1. Initiation of checkpointing at regular intervals is done by processes. The initial checkpoint is taken upon system initialization and initiated by $P_0$. The next checkpoint initiation is done by $P_1$ and so on and so forth.
2. Asynchronous communication has been assumed among the processes. Acknowledgement and time-out are part of the communication protocol.
3. A process is aware of the TMR group it belongs to and its role in that group.
4. Any AdhocNet routing algorithm may be used.

# 4. CHECKPOINTING

## 4.1 The Algorithm

The algorithm has a checkpoint initiator and uses explicit checkpoint synchronization messages. The initiator sends the initiation message to all others along with further information: number of messages sent to processes in the current checkpointing interval and number of messages received from processes in the current checkpointing interval. It must be mentioned here that the additional information regarding messages would not be sent during the initial checkpoint since it is taken just after the system has been initialized and hence it is assumed that communication among processes has not yet started. The information means that if $P_k$ has sent a total of two messages to $P_j$ in the current checkpointing interval, then $P_k$ would write 2 as number of messages and j as process id as part of the first information. Similarly if $P_j$ has indeed received all the two messages from $P_k$ it would write 2 as number of messages and k as process id as part of the second information. $P_j$ checks whether the total number of messages sent by $P_k$ matches with that received by $P_j$. If the answer is positive, $P_j$ takes the checkpoint. If not, then it waits for the unreceived message/s and takes the checkpoint after receiving it/them. During this time only those messages are received for which $P_j$ is waiting and any unwanted message is discarded [20]. The algorithm works as follows:

The initial checkpoint is taken after system initialization in lines 7-10 (for the initiator) and lines 14-16 (for other processes) in algorithm1. For any other checkpoint, the initiator first sends a "request for checkpoint" message followed by a message containing its status information for the current checkpointing interval (lines 11-12). Any other process on receiving the above (lines 16-17) sends its own status information to all other processes (line 18) and waits for receiving such information from the others (line 19). After it receives status information from others it goes on to check whether there is any message that has been sent by some other process to it but not yet received by it (lines 20-23). It waits to receive the said message/s and then takes the checkpoint (lines 24-25).

The variables used in the algorithm are described as follows:

| | |
|---|---|
| initiator: | pid of checkpoint initiator |
| check_index: | checkpoint sequence number |
| own_pid: | self process id |
| msg_type: | denotes a tag for identifying various kinds of messages: |
| | 0: checkpoint-request message |
| | 1: process-status-information message |
| | any other: computation message |
| mess_sent_to$_i$[j]: | an array; values of whose indices denote the number of messages that the concerned process has sent to (i.e. if the value in mess_sent_to$_i$[j] is n, this means that $P_i$ (concerned process) has sent n messages to $P_j$ in the current checkpointing interval) |
| mess_recd_fm$_i$[j]: | an array; values of whose indices denote the number of messages that the concerned process has received from (i.e. if the value in mess_recd_fm$_i$[j] is n, this means that $P_i$ (concerned process) has received n messages from $P_j$ in the current checkpointing interval) |

The subroutines used in the algorithm are as follows:

send, receive:              communication primitives
take_checkpoint:      saves process state
recv_sp:                      for executing the receive communication primitive with added
                                       logic like checking message type, message sequence number or
                                       even the sender etc
send_sp:                      for executing the send communication primitive with added logic
                                       like checking message type, message sequence number or even the
                                       receiver etc

The structure of the checkpointing algorithm is given below along with the line numbers mentioned in the leftmost column:

```
1    Procedure Checkpoint(P_i)
2    {
3    initiator := 0;
4   check_index := 0;
5   dest_id := -1;
6   if (initiator = own_pid)
7       if  (check_index = 0)
8           { msg_type = 0;
9               send_sp(msg_type, dest_id, check_index, seq_no);
10              take_checkpoint; }
11      else{msg_type:= 0; send_sp(msg_type,dest_id, check_index,seq_no);
12       msg_type := 1; send_sp(msg_type,dest_id, check_index, seq_no);}
13   else if (check_index = 0) {
14           recv_sp(recd_msg_type, send_id, recd_check_index, seq_no);
15           take_checkpoint; }
16       else{recv_sp(recd_msg_type,send_id, recd_check_index,seq_no);
17          recv_sp(recd_msg_type, send_id, recd_check_index, seq_no);
18          send_sp(recd_msg_type, dest_id, check_index, seq_no);
19          recv_sp(recd_msg_type, send_id, recd_check_index, seq_no);
20          for (i = 0; i<= n-1, i++)
21          for (j = 0; j <= n-1 , j++) {
22            if (i ≠ j) {
23              if (mess_sent_to_i[j] ≠ mess_recd_fm_j[i])
24              if (own_pid = j)
25              recv_sp(recd_msg_type,send_id,recd_check_index,seq_no);}
26          take_checkpoint;   }
27   check_index := check_index + 1; }
```

The algorithm recv_sp works as follows: In lines $2 - 6$ it receives only those messages whose checkpoint numbers equal to that of the receiver's checkpoint number and message sequence number matches the expected message sequence. In lines $7 - 9$ the algorithm receives the checkpoint initiation messages. In lines $10 - 12$ the algorithm receives messages from other processes containing corresponding status information.

```
1    Procedure recv_sp(mtype,pid, checkid, seqno)
2    {
3     If ((mtype <> 0) OR (mtype <>1))
```

```
4          { If  ((checkid = check_index) AND((seqno<= (mess_sent_to_i[j])
5                AND ((seqno >=      (mess_recd_fm_j[i] + 1))) )
6                receive(rmtype, pid, check_id, r_seq, recv_mess)  }
7      else  ( if ( mtype = 0)
8                if (checkid = 0)  receive(rmtype, checkid ) from P_0 ;
9                else receive(rmtype, checkid);
10   else (if ( ( mtype = 1)  AND (checkid <> 0) AND (pid = -1) )
11             for (k = 0; ((k<=n) AND (k <> own_pid)); k++)
12                receive(rmtype,checkid, mess_sent_to, mess_recd_fm);
13   }     }
```

The algorithm send_sp works as follows: In lines $2-4$ it sends computation messages and in lines $5-8$ it sends own status information to all others.

```
1     Procedure send_sp(mtype,pid, checkid, seqno)
2     {
3      if (((mtype <> 0) OR (mtype <>1)) AND (pid <> -1))
4           {send(mtype, pid, checkid,seqno, mess)   }
5        else
6           if ( (( mtype = 0)  OR (mtype = 1))
7                { for (k = 0; ((k<=n) AND (k <> own_pid)); k++)
8                send(mtype, checkid, mess_sent_to, mess_recd_fm);
9     }                      }
```

## 4.2  Brief Proof

Theorem:

The checkpoints taken by the algorithm form a consistent global checkpoint.
Proof: The theorem is proved by contradiction. Let the checkpoints form an inconsistent global checkpoint. Then there should be a checkpoint $CP_i^k$ that happens before [1] another checkpoint $CP_j^k$. This implies that (i) there is at least a message m sent by $P_i$ after $CP_i^k$ but received by $P_j$ before $CP_j^k$ and (ii) there is at least a message m sent by $P_i$ before $CP_i^k$ but received by $P_j$ after $CP_j^k$. This can be proved in the following way:

It must be mentioned here that, case ii stated above does not make $CP_i^k$ happen before $CP_j^k$. Hence it is not mandatory that messages recorded "sent" in $CP_i^k$ should also have to be recorded "received" in $CP_j^k$.

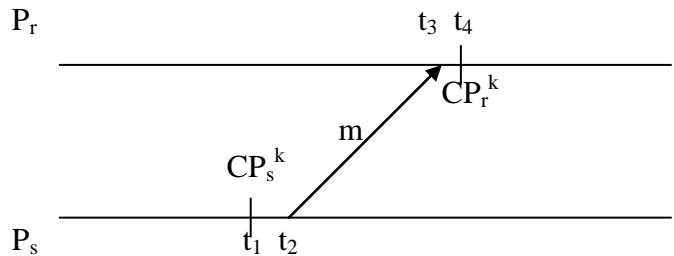

Figure 1. Message recorded received and not sent

Case (i): Let us consider figure 1 and a fault-free scenario where messages reach destinations correctly.

Assumptions:

1. Message m not recorded sent
2. Message m recorded received

The following scenario is observed:

i.  Message m is sent at $t_2$
ii. Message m is received at $t_3$
iii. Checkpoint $CP_s^k$ of $P_s$ is taken at $t_1$ ($t_1 < t_2$ by assumption 1)
iv. Checkpoint $CP_r^k$ of $P_r$ is taken at $t_4$ ($t_3 < t_4$ by assumption 2)
v.  Since $P_s$ takes checkpoint at $t_1$ (by assumption 1 and step iii)
    a. $P_s$ has reached line 26 of algorithm via lines 16-25.
    b. $P_s$ has checked its consistency with other (n-1) processes including $P_r$ in lines 18 – 25.
vi. In line 18 $P_s$ sends its status and $P_r$ receives it in line19 in $P_r$'s algorithm
    a. $P_r$ is in lines 19-25 and no discrepancies are noted.
    b. Therefore, $P_r$ reaches line 26 and hence takes checkpoint $CP_r^k$. (by iv) thereby violating assumption 2 and scenario ii and iv.
    c. Message m reaches $P_r$ and eventually gets rejected in lines 4-5 of procedure recv_sp of $P_r$ since m carries a later checkpoint index (by i and iii).
    d. vi (b, c) contradicts assumption 2.

Thus, there can not be any message m that is not recorded sent but recorded received in the same global checkpoint.

Alternative Proof:

With assumptions remaining the same, the following scenario is observed:

i.   Message m is sent at $t_2$
ii.  Message m is received at $t_3$
iii  Checkpoint $CP_s^k$ of $P_s$ is taken at $t_1$ ($t_1 < t_2$ by assumption 1)
iv.  Checkpoint $CP_r^k$ of $P_r$ is taken at $t_4$ ($t_3 < t_4$ by assumption 2)
v.   Assuming $P_r$ takes checkpoint at $t_4$ (by assumption 2 and step iv)
    a. $P_r$ has checked its consistency with other (n-1) processes including $P_s$ in lines 18 – 25 thereby confirming that all messages sent by $P_s$ have been received by $P_r$ and vice versa.
    b. $P_r$ has reached line 26 and taken checkpoint via lines 16-25.
    c. v (a, b) contradicts assumption 1.

Thus, there can not be any message m that is not recorded sent but recorded received in the same global checkpoint.

Case (ii) : Let us consider figure 2.

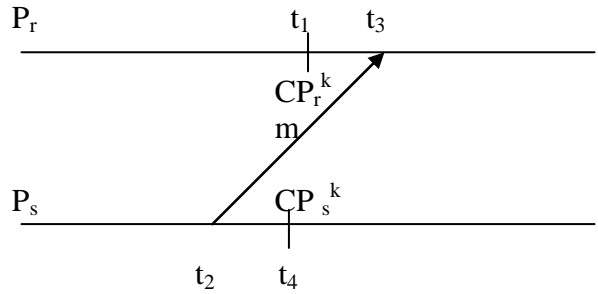

Figure 2. Message recorded sent and not received

Assumptions: 1. Message recorded sent. 2. Message not recorded received.
The following scenario is observed:

i.　　　　Message m is sent at $t_2$
ii.　　　　$P_s$ takes checkpoint at $t_4$ ($t_2 < t_4$ by assumption 1)
iii.　　　　$P_r$ takes checkpoint at $t_1$
iv.　　　　Message m is received at $t_3$ ($t_1 < t_3$ by assumption 2)
v.　　　　Assuming $P_s$ takes checkpoint at $t_4$
　　　　　a.　$P_s$ reaches line 26 (and records sending of m (by (ii))) via lines $16 - 25$.
　　　　　b.　$P_s$ has checked its consistency with other (n-1) processes including $P_r$ in lines $18 -$
　　　　　　　25.
vi.　　　Similarly, when $P_r$ takes checkpoint at $t_1$
　　　　　a.　$P_r$ reaches line 26 via lines $16 - 25$.
　　　　　b.　$P_r$ has checked its consistency with other (n-1) processes including $P_s$ in lines $18 -$
　　　　　　　25.
　　　　　c.　$P_r$ finds that message m from $P_s$ is yet to be received by it (by iv)
　　　　　d.　$P_r$ is in line 25 via lines $20 - 24$ until m is actually received in line 24.
　　　　　e.　$P_r$ can not reach line 26 and hence can not take checkpoint.
　　　　　f.　vi (e) contradicts assumption 2.

Hence, there can not be any message that is recorded "sent' but not recorded "received" in the
present checkpointing protocol.

# 5. RECOVERY

## 5.1 Approach to Recovery

Whenever consensus about the failure of a process is reached, it is also decided that processes
should rollback in order to restart from the last saved consistent state. Since not all processes are
dependent on the failed process in the concerned checkpointing interval so all of them need not
roll back. The processes that communicated with the failed process should roll back and they are
termed as being "directly" dependent on the failed process. Still there are others who have
communications with the directly dependent processes. Hence recovery of the directly dependent
processes would affect these "indirectly" dependent processes. So, they have to roll back also.
The task of finding whether a process is indirectly dependent on the faulty process has been taken up using several methods in the literature. The technique pursued here is described with the help of an example for better understanding.

Let us assume that in a system of 5 processes process2 is found to be faulty. The vectors used in the checkpointing algorithm that save (i) messages sent to and (ii) messages received from are sent by each process to all others. Let us consider figure 3 below and construct the above-said vectors for all the five processes.

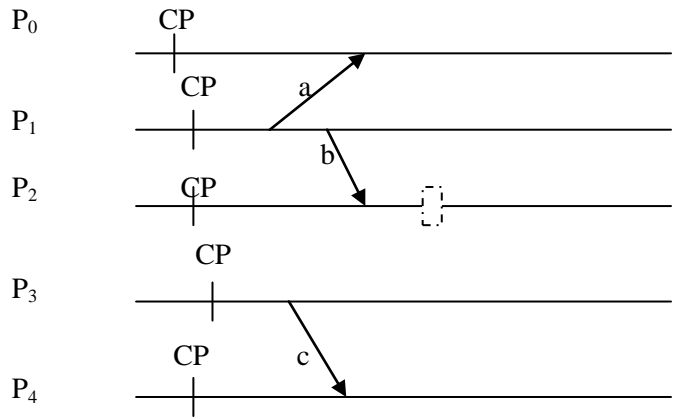

Figure 3. A scenario of process interactions via messages

The CP indicates the last consistent checkpoint of each process. ⌐⌐ represents the point where failure is detected. The entry "-1" is used to denote end.

| Process id | Message sent to (pid) | Message received from (pid) |
|---|---|---|
| 0 | -1 | 1, -1 |
| 1 | 0, 2, -1 | -1 |
| 2 | -1 | 1, -1 |
| 3 | 4, -1 | -1 |
| 4 | -1 | 3, -1 |

After the above vectors are available, each process builds the "sr" data structure (an array used in the Detect_Recovery algorithm) in a distributed fashion. In each process the "sr" array looks like the following array:

| Process id | Message sent to (pid)/ Message received from (pid) |
|---|---|
| 0 | 1, -1 |
| 1 | 0, 2, -1 |
| 2 | 1, -1 |
| 3 | 4, -1 |
| 4 | 3, -1 |

Let us now find out the dependency of each of the processes: (considering the faulty process id to be 2)

$P_0$:
- Searches own sr entry to find if 2 exists there.
- Since 2 is not there, so search the entry of sr[1] since 1 occurs in $P_0$'s sr-entry. $P_0$ keeps track that it has searched its own sr-entry.
- Gets 2 in sr[1].
- Concludes that $P_0$ has to roll back.

$P_1$:
- Searches own sr entry to find if 2 exists there.
- Since 2 is there, concludes that $P_1$ has to roll back.

$P_2$:
- Searches own sr entry to find if 2 exists there.
- Since 2 is not there, so search the entry of sr[1] since 1 occurs in $P_2$'s sr-entry. $P_2$ keeps track that it has searched its own sr-entry.
- Gets 2 in sr[1].
- Concludes that $P_2$ has to roll back.

$P_3$:
- Searches own sr entry to find if 2 exists there.
- Since 2 is not there, so search the entry of sr[4] since 4 occurs in $P_3$'s sr-entry. $P_3$ keeps track that it has searched its own sr-entry.
- Since 2 is not in sr[4], so search the entry of sr[3] since 3 occurs in $P_4$'s sr-entry. $P_3$ keeps track that it has searched sr-entry of 4.
- sr[4] contains only 3 whose sr-entry has already been searched and 2 was not in sr[3].
- Concludes that $P_3$ does not have to roll back.

$P_4$:
- Searches own sr entry to find if 2 exists there.
- Since 2 is not there, so search the entry of sr[3] since 3 occurs in $P_4$'s sr-entry. $P_4$ keeps track that it has searched its own sr-entry.
- sr[3] contains only 4 whose sr-entry has already been searched and 2 was not there.
- Concludes that $P_4$ does not have to roll back.

Data structures used in the Algorithm for detecting recovery:

sr[n][n]:  This array is constructed in each process after it gets the send-receive vectors of all other processes. This array denotes the pids of processes to/from which a particular process $P_i$ ($i <= n$) has sent/received message during the current checkpointing interval. If process $P_1$ has sent messages to processes $P_6$, $P_2$, $P_0$ and has received messages from $P_4$ and $P_3$, then sr[1][n] will contain the elements as mentioned below:

| 6 | 2 | 0 | 4 | 3 | -1 |
|---|---|---|---|---|----|

The −1 in sr[1][5] indicates that valid data for row 1 ends.

depends[]: This vector contains the process ids whose send-receive vectors have been checked by a process to find whether it is dependent on the failed process. The end of this vector is indicated by −1.

## 5.2 Algorithm for Detecting Recovery

```
1   Procedure Detect_recovery(Pᵢ)
2     {
3        k := 0;
4        flag, flag1 := F;
5        while (sr[ownpid][k] = pid_faulty)
6         { flag := T; recover(Pᵢ); }
7        k1, v1 := 0;
8        while (NOT flag1)
9         { key := sr[ownpid][k1];
10          if (key == -1)
11             flag1 := T;
12          else
13            { k2 := 0;
14              while ((sr[key][k2]<> pid_faulty) OR(sr[key][k2] <> -1))
15                 k2 := k2 + 1;
16              if(sr[key][k2] == pid_faulty)
17                 { flag := T; recover(Pᵢ,1); }
18              else
19                 { depends[v1] := key; v1 := v1 + 1;
20                   depends[v1] := -1; k1 := k1 + 1; }
21       flag2, flag3 := F; key1, k4 := 0;
22       if (flag1)
23        {
24          while (NOT flag2)
25           { k3 := 0;
26             while (NOT flag3)
27              { key2 := sr[depends[key1]][k4];
28                if ((key2 <> ownpid) OR (key2 <> -1))
29                 {while((sr[key2][k3]<>pid_faulty)OR(sr[key2][k3]<>-1))
30                     k3 := k3 + 1;
31                if (sr[key2][k3] == pid_faulty)
32                    { flag,flag2,flag3 := T; recover(Pᵢ,1); }
33                else
34                  { j := 0;
35                    while ((depends[j] <> key2) OR (depends[j] <> 1))
36                        j := j + 1;
37                    if (depends[j] == -1)
38                       { j := j - 1; depends[j] := key2;
39                         j := j + 1; depends[j] := -1; }
40                  k4 := 0; }
41              else { if (key2 == -1)
42                      { key1 := key1 + 1; flag3 := T; k4 := 0; }
43                    else { k4 := k4 + 1; flag3 := T; }
44             }
45          }
```

```
46          if (depends[key1] == -1) flag2 := T;
47       }
48    if (NOT flag)
49       recover (Pi,-1);
```

The concept of dependency is used in the above algorithm for recovery to minimize the number of nodes that roll back their computation. Only those nodes that have a dependency on the failed node since the latter node's last checkpoint is required to roll back to maintain global consistency. After the nodes roll back to their last saved consistent state, they have to retrace their computation that has been undone due to rollback. Types of messages that have to be handled are:

1. Orphan messages: This situation will arise when the sender rolls back to a state prior to sending while the receiver still has the record of its reception. However these messages can not arise because whenever sender $P_i$ rolls back, receiver $P_j$ also rolls back because by the above algorithm $P_j$ becomes dependent on $P_i$.

2. Lost messages: This situation will arise when the receiver rolls back to a state prior to reception of a message that is being still recorded as sent by the sender. However these messages can not arise because whenever receiver $P_i$ rolls back, sender $P_j$ also rolls back because by the above algorithm $P_j$ becomes dependent on $P_i$.

Since the above algorithm considers both the "send" as well as the "receive" vectors of a process in calculating dependency, so logging of messages by sender is not necessary as was the case in Prakash et. al [14].

## 6. WORKING OF ADHOCNET-BASED TMR

The above sections 3 and 4 describe the working of the checkpointing and the recovery protocols. This section describes the working of the TMR in AdocNet. Let us consider the following figure 4 that depicts an AdocNet and the various TMR groups it has. The network has 6 MHs with communication links as shown.

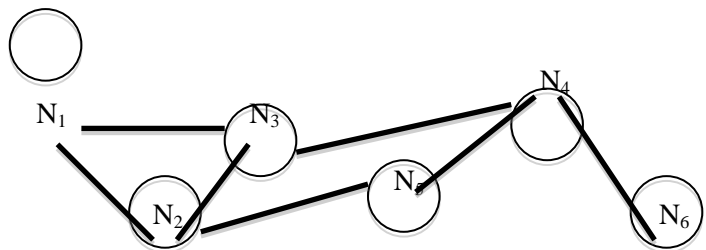

Figure 4. Example Wireless Adhoc Network

Let us consider the following wrt figure 4.

- TMR group1 or TMR1: $N_1$ (main), $N_2$ (primary) and $N_3$ (secondary)
- TMR group2 or TMR2: $N_3$ (main), $N_4$ (primary) and $N_5$ (secondary)
- TMR group3 or TMR3: $N_4$ (main), $N_5$ (primary) and $N_6$ (secondary)
  And so on.

Each MH knows its role in the group and also about the other two MHs belonging to its group. Hence, $N_3$ is aware that it acts as *secondary* in TMR1 and as *main* in TMR2. The responsibilities associated with the roles are different.

## 6.1 Checkpointing

The main MHs of each TMR are the only initiators in a group and can initiate checkpointing. Hence $N_1$ and $N_3$ and $N_4$ are the initiators here. According to the checkpointing algorithm described in the above section, checkpoint requests are received by all MHs and checkpoints taken accordingly depending on the activities in the current checkpointing interval. The MHs execute processes independently and the processes exchange messages frequently. The message exchange builds a dependence relation among them. A process executing on $N_1$ may also send message to $N_3$, belonging to the same TMR. Since any MH may take checkpoint in a particular checkpointing interval, the copies of its checkpoint are to be kept in the *main* and in the *primary* of that TMR. Whenever an MH is either *main* or *primary*, only one copy has to be sent to the other. But if an MH is *secondary*, then a copy each is sent to its *main* and *primary*. This may be an overhead in the network. However if the checkpointing interval can be chosen judiciously, this extra circulation of checkpoints would not be that much of an overhead. Whenever a new checkpoint is to be stored, the previous one is deleted in the corresponding *main* or *primary*. In case of failure of any one of the MHs in a TMR, that TMR reduces to Dual Modular Redundancy (DMR). In that case copies of checkpoints are with both the MHs in that group.

## 6.2 Recovery

Once the failed MH is identified (possibly after some time-out since message sending and non-receipt of acknowledgement), the processes in the system go to the *recovery* mode and exchange status information with each other. According to the recovery algorithm described above, a process is able to identify whether it should recover or not. It then proceeds to collect its checkpoint if it is *secondary*, otherwise the checkpoint is with itself only. Role change will happen to an MH if any other MH in its TMR is detected to have failed.

## 6.3 An Example Scenario

Suppose $N_3$ is detected to have failed. After subsequent status exchange, it is found that $N_2$ and $N_5$ are dependent on $N_3$. The latest checkpoints of $N_2$ and $N_5$ are with ($N_1$ and $N_2$) and ($N_4$ and $N_5$) respectively. Hence $N_2$ and $N_5$ have their checkpoints. $N_3$ was the *main* in TMR2 with $N_4$ (*primary*) and $N_5$ (*secondary*). Henceforth, $N_4$ becomes *main* and $N_5$ becomes *primary* in TMR2. Another important issue that needs to be considered in this changed scenario is that, henceforth $N_4$ will take up the role of $N_3$. Hence the MHs is the network may be made aware that the process running on $N_3$ would now execute on $N_4$. This is an additional task for $N_4$. However, generally, this should not pose any hindrance to the working scenario in the network.

## 7. CONCLUSION

The checkpointing algorithm proposed in this paper constructs consistent checkpoints in a distributed manner. Hence, forced checkpoints as well as useless checkpoints are never taken.

The checkpointing protocol described in the present work also eliminates the occurrences of both missing and orphan messages. Thus, each and every checkpoint taken by a process contributes to a consistent global snapshot and hence only the last global snapshot is required to be retained. The overhead of the present checkpointing protocol is the $\Theta(n^2)$ number of messages required during checkpointing (where n is the total number of processes). Though other algorithms have $\Theta(n)$ number of messages for the same but drawbacks like checkpoint commit time, failure of checkpoint coordinator, handling multiple checkpoint initiations are associated with them. Recovery of self is decided by each of the processes after collecting system-wide information. The dependence relation among the processes can be tracked on-line. A minimum number of processes is required to recover depending on their relation with the failed process.

Moreover this fault tolerance technique of checkpointing and recovery is based on TMR concept and that too in a wireless adhoc network. This paper proposes the approach towards obtaining fault tolerance using checkpointing and recovery on wireless adhoc network based TMR. The technique adopted is able to tolerate both the transient and permanent faults. The number of faults that can be tolerated is maximum two in each group of the TMR MHs in the wireless adhoc network.

This work does not consider node mobility in the adhoc network. However, the proposal can be extended to mobile ad hoc network.